\documentstyle[floats,twocolumn,prl,aps,psfig]{revtex} 

\newcommand\beq{\begin{equation}}
\newcommand\eeq{\end{equation}}
\newcommand\bea{\begin{eqnarray}}
\newcommand\eea{\end{eqnarray}}
\newcommand{\nonum}{\nonumber}

\begin{document}

\title{\Large Field theoretical study of a spin-1/2 ladder with unequal chain
exchanges}

\author{\bf Sujit Sarkar$^1$ and Diptiman Sen$^2$}
\address{\it $^1$ Max-Planck-Institute for the Physics of Complex Systems,
Noethnitzer Strasse 38, 01187 Dresden, Germany \\
$^2$ Centre for Theoretical Studies, Indian Institute of Science, 
Bangalore 560012, India}

\date{\today}

\maketitle

\begin{abstract}
We study the low-energy properties of a Heisenberg spin-1/2 zigzag ladder 
with different exchange constants on the two chains. Using a nonlinear 
$\sigma$-model field theory and abelian bosonization, we find that the 
excitations are gapless, with a finite spin wave velocity, if the values 
of the chain exchanges are small. If the chain exchanges are large, the 
system is gapped, and the energy spectra of the kink and antikink 
excitations are different from each other.
\end{abstract}

\vskip .4 true cm
\noindent PACS numbers: ~75.10.Jm, 75.10.Ee, 71.10.Hf


\section{Introduction}

For the last several decades, one-dimensional and quasi-one-dimensional 
quantum spin systems have been studied extensively due to their unusual 
properties. Experimentally, many such systems are known to present a wide
range of unusual properties, and a variety of analytical and numerical 
techniques exist for studying these systems theoretically.
The three observations which make the low-dimensional spin systems 
particularly intersting are, (i) Haldane's conjecture for one-dimensional 
antiferromagnetic spin systems \cite{hal,aff1}, (ii) the discovery of 
high-temperature superconductivity and its magnetic properties at low doping 
\cite{bed}, and (iii) the discovery of ladder materials \cite{joh,hir}.

In spin ladders, two or more one-dimensional spin chains interact
with each other. For ladders with the railroad geometry, it has been observed 
that spin-1/2 systems with an even number of legs are gapped, while systems 
with an odd number of chains have gapless excitations \cite{dag1,dag2,cha}. 
However, the frustrated zigzag ladder \cite{cas} 
shows gapless spin liquid state or the gapped dimer state, depending 
on the ratio of the exchanges of the rungs to the chains \cite{aff2}.
 
Another interesting kind of system is the spin-Peierls system such as 
$CuGe O_3$. The authors of Ref. \cite{cas} have explained the form of the 
ground state and the low-temperature thermodynamic properties of this sample 
by showing that there is spontaneous dimerization of the nearest neighbor 
interaction below a particular temperature. Their model includes a
dimerization in the nearest neighbor exchange coupling.

Relatively less is known about a spin ladder with asymmetry in the chains.
An extreme case of this situation, {\it i.e.}, when a chain is absent from 
the Fig. 1 (sawtooth chain), has been studied by two different groups 
\cite{naka,sen}. The ground state is like that of the Majumdar-Ghosh 
model \cite{maj}, except that the two kinds of low-energy excitations 
(kinks and antikinks) which interpolate between the two degenerate ground 
states have quite different excitation spectra.

In this paper, we study a two-chain ladder system
with unequal exchange constants of the chains, rather than a dimerization 
in the rungs. Very recently, Voit {\it et al} \cite{voit} have studied this 
problem. We will discuss their results below. The plan 
of the paper is as follows. In Sec. II, we will analyze the problem using
the nonlinear $\sigma$-model (NLSM) field theory. Sec.\ III will discuss the 
abelian bosonization approach.
 
\section{Nonlinear $\sigma$-model study}

In this section we study the NLSM field theory of our model. 
The schematic plot is shown in Fig. 1. The system can be viewed either
as a ladder with unequal exchanges on the two chains, or as a single
chain with unequal next-nearest neighbor exchanges. The Hamiltonian is
\bea
H &=& \sum_n ~[~ J_1 \vec{S}_n \cdot \vec{S}_{n+1} + J_2 (1 - (-1)^n ~
\delta) \vec{S}_{n} \cdot \vec{S}_{n+2} ~] ~, \nonum \\
&&
\label{ham}
\eea
where $n$ is the site index, $J_1 , J_2 \ge 0$, and $0 \le \delta \le 1$. 
If we view the system as two chains, then the lower (upper) chain 
contains the odd (even) numbered sites.

\begin{figure} 
\psfig{figure=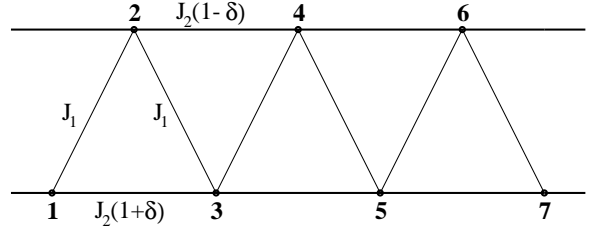,width=8cm,angle=-90} 
\caption{Schematic diagram of a spin ladder with unequal chain exchanges.} 
\label{fig1}
\end{figure}

Using the NLSM field theory, one can describe the 
low-energy and long-wavelength excitations. In the Neel phase, 
this is given by an $O(3)$ NLSM with a topological
term. Here we define two fields $\vec{\phi}_n $ and 
$\vec{l}_n$ as a linear combination of two spins as
\bea
\vec{\phi}_n &=& \frac{\vec{S}_{2n-1} - \vec{S}_{2n}}{2S} ~, \nonum \\
\vec{l}_n &=& \frac{\vec{S}_{2n-1} + \vec{S}_{2n}}{2a} ~,
\label{phil}
\eea
where $a$ is the lattice spacing. It can be easily checked that 
\bea
\vec{\phi}_n \cdot \vec{l}_n &=& 0 ~, \nonum \\
\vec{\phi}_n^2 & = & 1 + \frac{1}{S} - \frac{a^2 \vec{l}_n^2}{S^2} ~.
\label{cons}
\eea
Thus $\vec{\phi}_n$ becomes a unit vector in the large $S$ limit.
The unit cell of the classical ground state of the Neel phase is labeled by 
an integer $n$, and it contains the sites $2n-1$ and $2n$ respectively; the
length of a unit cell is $2a$.

The fields $\vec{\phi}_n$ and $\vec{l}_n$ satisfy the commutation relations
\beq
[\vec{l}_{ma}, \vec{\phi}_{nb}] = \frac{i}{2a} ~\delta_{mn}~
\sum_c ~\epsilon_{abc} ~\vec{\phi}_{nc} 
\eeq
where $m$ and $n$ are the unit cell labels, $a,b,c$ denote the
$x,y,z$ components of the field, and ${\epsilon}_{xyz}$ is the completely 
antisymmetric tensor with $\epsilon_{xyz}~= ~1$. This relation enables us
to write $\vec{l}_n = \vec{\phi}_n \times \vec{\Pi}_n$, where the
vector $\vec{\Pi}$ is canonically conjugate to $\vec{\phi}$ namely,
\beq
[\phi_{ma}, \Pi_{nb}]~= \frac{i}{2a} \delta_{mn} \delta_{ab} ~.
\eeq
To define the continuum limit of this theory, we introduce 
a spatial coordinate $x$ which is equal
to $2na$ at the location of the ${n}^{th}$ unit cell. Summations are then
replaced by integrals, {\it i.e.}, $\sum_n \rightarrow \int dx/(2a)$. 

Since $\vec{\phi}_n$ is a unit vector, $\dot{\vec{\phi}}$ and 
$\vec{\phi}'$ are orthogonal to $\vec{\phi}$.
In the low-energy and long-wavelength limit, the dominant
terms in the Hamiltonian are those which have second order space-time 
derivatives of $\vec{\phi}$ and first order derivatives of $\vec{l}$
(since $\vec{l}$ contains the the first order derivatives of $\vec{\phi}$) 
\cite{rao,sen2}. To get the continuum Hamiltonian, we expand the fields 
$\vec{\phi}_{n+1} = \vec{\phi} (x+2a)$ and $\vec{l}_{n+1} =\vec{l} (x+2a)$,
where $x=2na$, as 
\bea
\vec{\phi} (x + 2a) &=& \vec{\phi} (x) + 2 a \vec{\phi}' (x)
+ 2 a^2 \vec{\phi}'' (x) + \cdots ~, \nonum \\
\vec{l} (x + 2a) &=& \vec{l} (x) ~+ ~2a \vec{l}' (x) + \cdots ~.
\label{tayl}
\eea
Using Eqs. (\ref{phil}), (\ref{cons}), and (\ref{tayl}), we obtain the 
Hamiltonian
\beq
H = \int dx~ [~\frac{c g^2}{2}~ ( \vec{l} ~+ ~\frac{\theta}{4 \pi} 
\vec{\phi}' )^2 ~+~\frac{c}{2g^2} \vec{\phi}'^2 ~] ~,
\label{hamnlsm}
\eeq
where
\bea
c & = & 2 J_1 S a \sqrt{1 - 4 J_2 /J_1} ~, \nonum \\
g^2 & = & \frac{2}{S \sqrt{1~-~4 {J_2}/J_1}} ~, \nonum \\
\theta & = & 2 \pi S ~. 
\label{coeff}
\eea
Note that the values of $c, g^2$ and $\theta$ turn out to be independent of 
$\delta$ in this approach. Further, this NLSM is valid only if $J_2 /J_1 < 
1/4$. For $J_2 /J_1 > 1/4$, a different NLSM is required.

One can find the energy-momentum dispersion relation of the form 
$\omega~= c |k|$, where c is the spin wave velocity, by considering small 
fluctuations around $\vec{\phi} = (0,0,1)$, and expanding the Hamiltonian in 
(\ref{hamnlsm}) to second order in those fluctuations.
Similarly one can find the strength of the interaction between the spin 
waves, $g^2$, by expanding the Hamiltonian to fourth order in the fluctuations.

{}From Eq. (\ref{coeff}), we see that the coefficient of the 
topological term $\theta =\pi$ for a spin-1/2
system. This implies that there is no gap in the low-energy excitation 
spectrum. This result is different from that of the zigzag ladder with a
dimerization in the nearest neighbor interaction, {\it i.e.}, with a term
like $J_1 (-1)^n {\vec S}_n \cdot {\vec S}_{n+1}$. In that case
$\theta~= 2 \pi S ~(1 - \delta)$ is different from $\pi$ for a spin-1/2 
system \cite{sen2,sie}, and the the low-energy excitations are gapped.

\section{Abelian bosonization study}

In this section we study the low-energy excitations spectrum
using abelian bosonization. We express our Hamiltonian as the following sum,
\beq
H~= H_1~ +~ H_2 ~+~ H_{2\delta} ~,
\eeq
where 
\bea
H_1 & = & J_1 ~\sum_{n}~ \vec{S}_n \cdot \vec{S}_{n+1} ~, \nonum \\
H_2 & = & J_2 ~\sum_{n}~ \vec{S}_n \cdot \vec{S}_{n+2} ~, \nonum \\
H_{2\delta} & = & - J_2 ~\delta ~\sum_{n}~ (-1)^n ~\vec{S}_n \cdot 
\vec{S}_{n+2} ~.
\label{hams}
\eea
We can convert this Hamiltonian to a Hamiltonian of spinless fermions 
through the Jordan-Wigner transformation. The relations between the spin and
the electron creation and annihilation operators are,
\bea
S_n^z & = & \psi_n^{\dagger} \psi_n - 1/2 ~, \nonum \\
S_n^- & = & (-1)^n \psi_n ~\exp [i \pi \sum_{j=-\infty}^{n-1} n_j]~, \nonum \\
S_n^+ & = & (-1)^n \psi_n^{\dagger} ~\exp [-i \pi \sum_{j=-\infty}^{n-1} n_j]~,
\eea
where $n_j = \psi_j^{\dagger} \psi_j$ is the fermion number at site $j$.
The Hamiltonians in (\ref{hams}) then become
\bea
H_1~&=& - \frac{J_1}{2} ~\sum_n ~(\psi_{n+1}^{\dagger} \psi_n + 
\psi_n^{\dagger} \psi_{n+1}) \nonum \\
&& + J_1 ~\sum_n ~(\psi_n^{\dagger} \psi_n - 1/2) (\psi_{n+1}^{\dagger} 
\psi_{n+1} - 1/2) ~,
\label{h1fer}
\eea
\bea
H_2~&=& J_2 ~\sum_n ~( \psi_{n+2}^{\dagger} \psi_n + {\rm h.c.})
(\psi_{n+1}^{\dagger} \psi_{n+1} - 1/2) \nonum \\
&& +~ J_2 ~\sum_n ~(\psi_n^{\dagger} \psi_n - 1/2)
(\psi_{n+2}^{\dagger} \psi_{n+2} - 1/2) ~,
\label{h2fer}
\eea
\bea
H_{2\delta} &=& - J_2 \delta \sum_n (-1)^n (\psi_{n+2}^{\dagger} 
\psi_n + {\rm h.c.}) (\psi_{n+1}^{\dagger} \psi_{n+1} - 1/2) \nonum \\ 
&& - J_2 \delta \sum_n (-1)^n (\psi_n^{\dagger} \psi_n - 1/2)
(\psi_{n+2}^{\dagger} \psi_{n+2} - 1/2) . \nonum \\
&&
\label{h2dfer}
\eea
We will assume below that $J_1 >> J_2$. Since there is no applied 
magnetic field, the two Fermi points occur at $ k_F = \pm \pi/2$.
We can linearize the energy spectrum around these Fermi points, and 
express the lattice operators in terms of two continuum fields $R$ and $L$
which vary slowly on the scale of a lattice spacing, 
\beq
\psi_n ~=~ {\sqrt a} ~[~ i^n R(n)~+~(-i)^n L(n) ~] ~,
\label{rl}
\eeq
where $R$ and $L$ are describe the second-quantized fields of right- and 
left-moving fermions respectively. Now we bosonize our Hamiltonian following 
the standard procedure. The basic relations used to obtained the bosonized 
Hamiltonian are as follows \cite{gog}.
\beq
S^{z} (x)~= ~a ~[ ~\rho (x)~+~ (-1)^j ~ M(x) ~] ~,
\label{szbos}
\eeq
where the fermion density $\rho (x)= :R^{\dagger} (x) R (x): + : L^{\dagger} 
(x) L(x):$, and the mass operator $M(x)= :R^{\dagger} (x) L(x): + : 
L^{\dagger} (x) R(x): $. (The double dots denote normal ordering). 
The bosonized expressions for $\rho$ and $M$ are given by
\bea
\rho (x) &=& ~-~ \frac{1}{\sqrt \pi} ~\partial_x \phi (x) ~, \nonum \\
M(x) &=& ~\frac{1}{\pi a} ~\cos (2 {\sqrt \pi} \phi (x)) ~.
\eea

The bosonized version of $H_1$ is known to be
\bea
H_1 &=& \int dx ~[~ \frac{v_1 K}{2} ~\Pi^2 ~+~ \frac{v_1}{2K} ~(\partial_x 
\phi)^2 \nonum \\
&& ~~~~~~~~ + \frac{v_1}{(\pi a)^2} :\cos (4 {\sqrt \pi} \phi ): ~] ~,
\label{h1bos}
\eea
where $v_1 =\pi J_1 a/2$ is the spin wave velocity, and $K=1/2$ is the 
bosonization interaction parameter for the isotropic spin-1/2 antiferromagnet.
The last term in (\ref{h1bos}) is marginal, and is known to have no effect 
at long distances in the sense of the renormalization group (RG) \cite{gog}. 
Using Eqs. (\ref{h2fer}), (\ref{rl}), and (\ref{szbos}), we get the following 
expression for $H_2$, where we have ignored terms of the order of $a^4$ and 
higher, 
\bea
H_2 &=& J_2 a^2 ~\sum_n ~ [ - ~(\rho_{n+1} -(-1)^n M_{n+1}) \times \nonum \\
&& ~~~~~~~~ (\rho_n + \rho_{n+2} + (-1)^n M_n + (-1)^n M_{n+2}) \nonum \\ 
&& ~~~ + (\rho_n + (-1)^n M_{n}) ({\rho}_{n+2} + ~(-1)^n M_{n+2})].
\eea
To derive this expression, we have used Taylor expressions such as
\beq
R(n+2) ~=~ R(n) ~+~ 2aR'(n) ~+~ 2a^2 R''(n) ~+~ \cdots 
\eeq
to write
\bea
&& R^{\dagger}(n+2) R(n) ~+~ R^{\dagger}(n) R(n+2) \nonum \\
&& ~=~ R^{\dagger}(n+2) R(n+2) ~ +~ R^{\dagger}(n) R(n) ~+~ O(a^2) ~.
\eea
On keeping only the terms which do not oscillate as $(-1)^n$ (which would 
give zero in the continuum limit) and then expressing the operators
$\rho$ and $M$ in the bosonic language, the above expression becomes
\beq
H_2 ~= v_2 ~ \int dx ~[ -\frac{1}{\pi} ({\partial}_x \phi)^2
+ \frac{3}{2 (\pi a)^2} \cos (4 {\sqrt \pi} \phi )],
\label{h2bos}
\eeq
where $v_2 = J_2 a$. Both the terms in (\ref{h2bos}) have scaling dimension 
2, and they are marginal. It is known that they have no effect in the RG 
sense as long as $J_2 /J_1 < 0.241$ \cite{oka}.
Finally, one can find the bosonized expression for $H_{2\delta}$. 
To begin with, the nonoscillatory part of the Hamiltonian is given by
\bea
H_{2\delta} &=& - J_2 \delta a^2 \sum_n ~[\rho_n M_{n+2} + M_n \rho_{n+2} - 
\nonum \\ 
&& ~~~~ \rho_{n+1} (M_n + M_{n+2}) + M_{n+1} (\rho_n + \rho_{n+2})].
\label{h2dbos}
\eea
Now we perform an operator product expansion of the above Hamiltonian. In the
limit $z \rightarrow w$, we can use the following expansion \cite{nag,eft},
\bea
\partial_z \phi (z) :e^{i \beta \phi (w)}: &=& - \frac{i\beta}{z-w} :e^{i 
\beta \phi (w)}: \nonum \\
&& + : \partial_z \phi (z) e^{i \beta \phi (z)}: ~
\label{ope}
\eea
for $K=1/2$. The second term in (\ref{ope}) is a total derivative, and its 
contribution will therefore vanish in the Hamiltonian where it appears inside 
an integral over all $x$. From the first term in Eq. (\ref{ope}), we see that
the various terms in Eq. (\ref{h2dbos}) cancel each other in pairs which have 
$z-w = \pm a$ or $\pm 2a$. If this cancellation had not occurred, the bosonized
version of $H_{2\delta}$ would have been proportional to the 
operator $\cos (2 {\sqrt \pi} \phi)$, which has scaling dimension 1 and
would therefore have been relevant. However, due to the cancellation,
$H_{2\delta}$ contains no relevant operators with scaling dimension less than 
2. Thus the system continues to remain gapless (and lies in the same 
spin-liquid phase as the model described by the nearest neighbor Hamiltonian 
$H_1$) even if $\delta \ne 0$, provided that $J_2 << J_1$. This is in 
contrast to a dimerization in $J_1$; in that case abelian bosonization 
correctly produces a relevant term which leads to a gapped phase.

Recently Voit {\it et al} \cite{voit} have studied the same model as ours 
using abelian bosonization followed by a perturbative 
renormalization group analysis. They claim that the perturbation due to the 
unequal chain exchanges ($J_2 \delta$ terms) is relevant but that it does not
lead to a gapped phase; instead, they argue that it leads to a different 
fixed point where the system is gapless and has a vanishing spin velocity.
In our study, we have shown that due to some cancellations, the $J_2 \delta$ 
terms do {\it not} lead to any relevant terms in the continuum theory,
and therefore that the system remains gapless if $J_2$ is small.

Finally, let us discuss the low-energy excitations of this model when $J_2$ 
becomes larger. In particular, we find that these are given by kinks and 
antikinks when $J_2 = J_1 /2$, for all values of $\delta$ lying in the range 
$[0,1]$. (The Majumdar-Ghosh model is a special case of this where 
$\delta =0$). Then the Hamiltonian in (\ref{ham}) can be written, up to a 
constant, as the sum
\bea
H ~=~ \frac{J_1}{4} ~\sum_n ~[ && (1+\delta )({\vec S}_{2n-1} +{\vec S}_{2n} +
{\vec S}_{2n+1})^2 \nonum \\
&& + (1-\delta )({\vec S}_{2n} +{\vec S}_{2n+1} + {\vec S}_{2n+2})^2] ~.
\eea
Hence the ground state is given by a configuration in which the total spin 
of each triangle is 1/2. Since this can be done either by forming a singlet 
with the pair of spins $(2n-1,2n)$ for all values of $n$, or by forming a 
singlet with the pair of spins $(2n,2n+1)$
for all values of $n$, we see that the ground state is doubly degenerate.
Let us denote these two ground states by A and B respectively.
The lowest-energy excitations (kink and antikink) are formed by interpolating
between these two ground states. The kink has the ground state A on the 
left and the ground state B on the right, while the antikink
has the ground state B on the left and the ground state A on the right.
For general values of $\delta$, we find that the kink
and antikink dispersion are nondegenerate in contrast with the Majumdar-Ghosh
model \cite{naka,sen}. A simple variational calculations gives the
kink and antikink dispersions to be $J_1 (1-\delta ) (5+4 \cos k)/8$ and
$J_1 (1+\delta ) (5+4 \cos k)/8$ respectively. Hence the
minimum gap for the kink and antikink excitations are $J_1 (1 - \delta)/8$ 
and $J_1 (1 + \delta)/8$ respectively, occurring at $k=\pi$. This result 
has been also obtained by the Voit {\it et al} \cite{voit}.

To summarize, we have studied the low-lying excitations of a 
zigzag ladder with unequal chain exchanges. Both the NLSM 
and abelian bosonization show that the system remains gapless if $J_2 << J_1$.
We have also shown that the system is gapped (with two degenerate ground
states) for $J_2 = J_1 /2$. It would be interesting to use numerical 
techniques like the density-matrix renormalization group \cite{pat}, to study 
the complete phase diagram of the ground state as a function of the two 
parameters $J_2 /J_1$ and $\delta$.

We (SS and DS) would like to thank Marco Ameduri for useful discussions. DS
would like to thank Bruce Normand for helpful comments.

\end{document}